# Practices in source code sharing in astrophysics

Lior Shamir[1], John F. Wallin[2], Alice Allen[3], Bruce Berriman[4], Peter Teuben[5], Robert J. Nemiroff[6], Jessica Mink[7], Robert J. Hanisch[8], Kimberly DuPrie[3]

1. Lawrence Technological University
2. Middle Tennessee State University
3. ASCL
4. Infrared Processing and Analysis Center, California Institute of Technology
5. University of Maryland
6. Michigan Technological University
7. Harvard-Smithsonian Center for Astrophysics
8. Space Telescope Science Institute

**Abstract**

While software and algorithms have become increasingly important in astronomy, the majority of authors who publish computational astronomy research do not share the source code they develop, making it difficult to replicate and reuse the work. In this paper we discuss the importance of sharing scientific source code with the entire astrophysics community, and propose that journals require authors to make their code publicly available when a paper is published. That is, we suggest that a paper that involves a computer program not be accepted for publication unless the source code becomes publicly available. The adoption of such a policy by editors, editorial boards, and reviewers will improve the ability to replicate scientific results, and will also make the computational astronomy methods more available to other researchers who wish to apply them to their data.

1. Introduction

The rapid advances in information technology and the increasing importance of digital sky surveys generating massive astronomical databases have reinforced the need for computational methods that can manage, store, process, analyze, and mine astronomical data, and the importance of software has become a fact of life (Weiner et al., 2009). *Astronomy and Computing* addresses the important bridge between information science and astronomy, and is an outlet for innovation in computational methodology for astrophysics.

However, method papers are of limited use to the astrophysics community unless the code and software that accompany the algorithms can be tested and validated by interested investigators and applied to their data. Furthermore, experimental results that were produced by using computer programs are often difficult to replicate without the source code used in the experiment. Unfortunately, scientific papers are often not provided with useable software for various reasons ranging from resistance to revealing code that researchers believe is not clean enough for sharing (Barnes, 2010), to job security and intimidation by potential requests for support by those who wish to use it (Teuben et al., 2012) and lack of time and resources to support these requests.

While many within the community already agree that source code should be shared, the attempts of the Astrophysics Source Code Library (ASCL, http://ascl.net) to make scientific source codes more discoverable to the public reveal a different, troubling picture. The editor of ASCL searches through recently published papers and asks authors of relevant papers whether their code is available for download. From that effort we learn that ~66% of the requests remain unanswered, and ~20% of the authors reply but indicate that the code will not be shared. Merely ~13% of the authors contacted by ASCL editors agree to make their source code publicly available. Typical reasons for not sharing the computer programs include "the code is not open source", "there is no documentation so that code is not released", or "an announcement will be made when the code is available". When ~87% of the authors refuse to share their code, and 66% do not believe that a request for sharing code justifies a reply, it appears that the general agreement often expressed at relevant conferences that source codes should be shared (Allen et al., 2012c; Allen et al., 2013) does not coincide with the way most researchers act when their own code becomes the subject of the discussion. In this paper we briefly discuss the need to make source code publicly available, and propose simple and inexpensive ways to achieve the goal of integrating the source code in the process of scholarly communication in astrophysics.

## 2. Advantages of sharing scientific source code

Sharing source code used in a scientific paper means that the code of a computational method that was used to produce the results is accessible to the public via the internet. That ensures that the public can read, validate, and criticize scientific source code, replicate the results described in the paper, and apply, refine, and extend the methodology described in the paper to their own data without the need to take the labor intensive task of re-implementing the algorithms. Authors of code described in a scientific paper can release it under the software license of their choice, as long as it allows free use of the software at least for replication of the results and for non-commercial scientific research. In any case, authors can retain intellectual property rights to the code, and are not required to transfer them to the publisher of the journal. The authors of the source code can also specify the acknowledgments, if any, that they request from those who wish to use the code.

As mentioned in Section 1, not all scientific software developers are willing to share the code that they describe in peer-reviewed papers. One of the important advantages of releasing source code is that it allows replication of the results, which is a key concept in science (Aristotle, 350 BC). Without access to the original code used in an experiment, replication will require re-developing the code, a task that in many cases involves significant labor, but even after completion of the development differences of the results can be attributed to errors or differences in the code. Releasing the original code allows easier replication of the results and ensures they are not vulnerable to coding errors or differences in the implementation of the algorithms.

Releasing the source code can also help interested readers to better understand the methodology. Indeed, Ince, Hatton and Graham-Cumming (2012) have argued that code should be shared regardless of its perceived quality, because dissection of code by a broad community is requisite to understanding its operation and validating science results. Furthermore, as astronomy has become dependent on computational methods that can automatically process large sets of data, data analysis methods published in scientific journals are in increasing demand. If a paper that describes a computational method is accepted for publication, the reviewers and editors who recommended publication believe that it will be of interest to others, otherwise they would have not accepted the paper. While researchers interested in the method can implement it based on the description in the paper, software development can become a labor-intensive task that requires significant resources. A few sentences in the methodology section of a paper qualitatively describing an algorithm cannot hope to capture the complexities of even a short thousand-line code. Most descriptions of algorithms are ill defined because of the need for brevity in scientific papers. The only true documentation of the analysis performed by software is the software itself. In addition, re-developing the code puts additional pressure on the already limited research resources, and will clearly not lead to optimal scientific return (Weiner et al., 2009).

Sharing of scientific software can significantly increase the impact of the published work by providing its target audience with the ability to easily test and use the method. For instance, it is reasonable to assume that methods such as SExtractor (Bertin & Arnouts, 1996) and GALFIT (Peng et al., 2002), which were used in thousands of studies according to ADS, would be of much lower impact if they were not released as software packages, and if every investigator interested in using them needed to implement the algorithm from scratch. Making the software publicly available makes it easy to test the methodology and evaluate its performance on any dataset.

Another advantage of sharing source code is that it allows comparing the performance of different methods for the purpose of testing how each method performs on given data. In the field of computer science it is common to use benchmarks -- public datasets that are used by algorithm developers to compare the performance of computational methods (Phillips, 2000; Samaria & Harter, 1994; Kussel & Baidyk, 2004; Fei-Fei, Fergus & Perona, 2006). The use of public benchmarks allows the comparison of the performance of different methods using the same data without the need to share the code. That is, a researcher can test a new method by applying it to the benchmark, and compare the results with performance figures provided in previously published papers describing methods tested with the same data. This approach has been used in thousands of scientific papers, but has also been criticized for leading to biased results and to methods that excel in analysis of specific datasets, but the performance evaluation cannot be generalized to the problem these benchmarks aim to help solve (Pinto, Cox & DiCarlo, 2008; Pinto, DiCarlo & Cox, 2009; Shamir, 2008; Shamir, 2011; Jenkins & Burton, 2008). Sharing the source code allows testing and comparing the performance of methods on different datasets, which can be the actual data that the researcher attempts to process, without being bound to a specific benchmark. Clearly, it also helps to validate the published performance figures by replicating the results. An example of the use of public availability of computer programs in astrophysics is the comparison of source extraction methods, which was made possible by the authors of the tested methods who made their software publicly available (Annunziatella et al., 2012; Huynh et al., 2012).

## 3. Practices to encourage source code sharing

As discussed in Section 2, sharing source code is of high value to astronomy, and can be achieved with very little sacrifice of research resources as will be described in Section 4. However, as discussed in Section 1, the actual current code sharing practices are not in agreement with the critical need to communicate the software and share it with the entire astrophysics community. To close that gap, we propose that a paper should not be published in a scientific journal unless the computer programs described in it are made publicly available. Making the code clean and well documented, preparing tutorials, synopsis of the application programmer interface (API), readme files, and other materials that make the code more useful require significant work, and therefore authors may be reluctant to release their code. To encourage the release of source code without additional pressure on the author's resources, we suggest that the code does not have to be perfectly clean or well documented (Barnes, 2010), but as part of the research it should be made accessible to the readers when the paper is published. Obviously, this policy of sharing source code should apply only to software developed by the authors and described in the paper, and not to common proprietary software tools (e.g., Microsoft Excel) that the authors do not have the ability or authority to share.

The importance of developing scientific software and making it publicly available has not yet been elevated to the level of publishing peer-reviewed papers for career development decisions such as promotion and tenure. Therefore, the requirement to make the code available when the paper is published will provide a tool for encouraging researchers to share their computer programs, and will eventually imprint the practice of sharing source code in the common practices of scientific communication. We argue that in order to effectively advance the sharing of software and source code all journals should make an explicit requirement that the code and software of the method described in a paper should become available to the scientific community upon publication, except in limited cases for compelling reasons such as national security concerns.

Some funding agencies such as the National Science Foundation require funded projects to describe how the outcome of the project, including the software, will be shared with the public at the time the proposal for the project is evaluated and the funding decision is made. However, the ability to replicate a scientific experiment should be independent of the mechanism by which the experiment was funded, and therefore journals cannot rely on the agreements between the investigators and their funding agencies. Moreover, funding agencies often do not monitor funded projects at the resolution of a specific paper at the time the paper is published, and therefore leaving the responsibility for sharing computer programs to funding agencies has been an ineffective mechanism for code release. For these reasons we argue that the availability of source codes should be part of the scientific communication process handled by the journal in which the paper is published.

The code shared with the public should include the implementation of algorithms and non-trivial data analysis described in the paper, as well as other pieces of code that were developed specifically for the study and cannot be re-written without investing a considerable programming effort. Trivial scripts are normally of less interest to the readers, and making such scripts public is not essential.

The requirement for making the software available when the paper describing it is published is not unique. For instance, *Bioinformatics*, one of the premier journals in its field, asks the authors upon submission to describe how the software can be accessed, and the URL that leads to the source code is specified in the title page of the paper. An example of a weaker requirement is the source code sharing policy of the journal *Annals of Internal Medicine*, for which authors are required to specify whether they will be willing to share the source code, data, and protocols they developed and used in the study after the paper is published (Laine et al., 2007). While authors can choose not to share any of the materials required for replication of the experiments, the advantage of this approach is that the refusal is public (Goodman, 2010). The journal *Biostatistics* has established a minimum standard for reproducibility of scientific results, and standards for submission of code and instructions of how it can be run against the data can be submitted with the paper (Peng, 2009). Authors who wish to meet these criteria submit their code and data to the associate editor for reproducibility (AER), who compiles the code according to the instructions of the authors and replicates the experiment before the paper is published (Peng, 2009).

The approach taken by *Biostatistics* requires an associate editor who needs to invest significant labor in replicating the experiments, and thus adds pressure on the review and publication process. Since the associate editor for reproducibility cannot be an expert in all studies published in the journal, in many cases the AER will not be able to effectively criticize the experiment and obtain deep understanding of its context, and therefore the reproducibility will come down to merely verifying that the output of the code processing the data on the authors' computer is similar to the output of the same code and data processed on the computer of the AER (Keiding, 2009). Moreover, in many astrophysics studies the datasets are extremely large and cannot be easily submitted, and the experiments require substantial computing power. A simple example is SDSS data release 7 (Abazajian et al., 2009), which was compiled using complex pieces of code and extremely large raw data, making it impractical for an editor to replicate the data release. Another difference between astrophysics and biostatistics is that while code development in statistics is dominated by a single programming language (the R language), astrophysicists use a variety of programming languages, compilers, and software development tools, and the associate editor for reproducibility might not have knowledge of or access to some of these tools, making the task of reproducibility more difficult. The option to satisfy the reproducibility criteria in *Biostatistics* is currently limited to software developed using the R programming language. In *Biostatistics,* publishing a paper that meets the "reproducibility" standard is optional, and authors are not required to submit the code and data unless they choose to publish their paper as a "reproducible" paper (Peng, 2009), so that papers based on source code developed in other programming languages are still valid for publication.

The variety of programming languages and software development tools used in astronomy, the large datasets, and the substantial computing resources often used in astronomical data analysis might make it challenging for editors to reproduce the results of all papers considered for publication. Also, hiring associate editors for reproducibility by all relevant journals will add financial burden and will probably delay the implementation of making scientific source codes publicly available.

Allowing the authors to choose whether they wish to make an experiment reproducible as done by *Biostatistics* has the downside of many authors selecting not to make their experiment replicable, since as described earlier in this section sharing the code and data has merely little value to the authors after the paper is published. The policy of *Annals of Internal Medicine* making the author's statement about reproducibility public provides some information about the willingness of authors to make their results reproducible. Out of 76 original research papers published in the journal in 2012, the authors of merely 15 papers declared that they would be willing to provide the statistical code, data, and protocols without restrictions. The authors of two paper indicated they would be willing to share their code and data but not the protocols, and nine more stated they would share the code and data subjected to the approval of the authors or after establishing collaboration between the authors and the person requesting the data. Only three papers that were published in 2012 made the source code available on-line, none of the authors of these papers were willing to provide also the data and the protocols without restrictions. Therefore, none of the papers published in *Annals of Internal Medicine* in 2012 can be replicated without the consent of the authors. The experience of ASCL described in Section 1 shows that the situation is not unique to *Annals of Internal Medicine*, and it is reasonable to assume that if other journals adopted the policy used by *Annals of Internal Medicine* the findings would be similar. The encouraging observation is that out of the 76 papers, 52 indicated that they would be willing to share the statistical source code if contacted by email. However, such communications are not public, and the experience of ASCL shows that contacting researchers by email after a paper is published to request the download site for the source code is in most cases not fruitful.

We propose that journals in astronomy and astrophysics should not require the reproduction of the results by the editor as a prerequisite for accepting a paper for publication. However, we urge that before a paper is published, the source code used in the experiments described in the paper be made publicly available. A simple form of implementing this practice is by adding a question to the report filled by the reviewer of the paper, which is whether the source code is publicly available. If the answer of the reviewer is "no", the paper should not be accepted for publication until the authors make their software publicly available. Exceptions can be made in cases where the code cannot be legally released due to national security concerns, or other compelling reasons that may be approved in the discretion of the editor. The source code does not have to be fully documented or well organized, but it should be accessible to anyone interested in reading or compiling it.

The downside of this approach is that there is just minimal inspection of the code before the paper is published, and there is no verification that the code released by the author indeed produces the results described in the paper. Other investigators using the software would have little or no support and need to understand the software on their own. However, the code becomes accessible and the experiment can be replicated by any reader who is willing to spend the time needed to understand the code. This approach will require authors to provide the same code used in the experiment, and the methodologies would be subject to analysis, criticism and replication after it is published. Without having access to the code, such analysis is extremely difficult. As described in Section 1, the experience of ASCL shows that some authors are willing to share their source codes while some are not, but we have not encountered cases in which authors were willing to share source code, but that code was not the code discussed in their paper.

The approach can be considered an impermanent solution which ultimately may be replaced by an editorial process that replicates all results. We believe that the simpler mechanism described in this section is more practical, as its implementation is immediate and inexpensive. It should also be noted that editors and reviewers do not normally replicate the results published in scientific papers, but ensure that there is sufficient details that allow repeating the experiment and reproducing the results.

Ultimately, astronomy and many other disciplines may be best served by repositories that link papers to codes and that allow readers to run the code on remote servers. The development and maintenance of such repositories will be expensive, yet the RunMyCode (http://runmycode.org) service, which has been live since 2012, has made considerable progress in showing how such a repository may operate (Stodden et al. 2012). Data and code are stored and executed on cloud platforms, and run from a web page linked from journal papers. While the service currently supports only a few languages and lacks a sustainability model, it is a highly promising first step in what may be the "next generation" repository.

4. Methods of sharing code in astrophysics

Perhaps the easiest way of sharing software is by simply uploading the files to the personal web site of one of the authors. The major weakness of doing so is that the code is then not easily discoverable. Another weaknesses of using personal web pages is the possibility that the web site will be shortlived, and that the files will disappear when the personal web site is closed (for instance, when the author of the code retires or moves to another institute). Another way of sharing source code is through code directories such as SkySoft, Astro-CodeWiki, and AstroShare (Shortridge, 2009). On-line services such as Github also offer tools for distributing, sharing, and archiving software. These services make the code more sustainable, but the codes are not indexed by a central system, and therefore using these services alone makes it difficult to index and search for computer programs.

The Astrophysics Source Code Library (ASCL) is an on-line index of source codes and programs used in astrophysics (Allen et al., 2012a; Allen et al., 2012b). Software indexed by ASCL can be uploaded to the ASCL web site, but can also be hosted using services outside ASCL such as Github. All source codes in ASCL are indexed by ADS so that they can be found in ADS queries, and also be cited by other papers such that the citations are counted by ADS. Currently, the editor of ASCL actively seeks for astrophysics programs published in peer-reviewed journals and makes them available to the community.

In some cases code can be complex, include very many files, and require numerous libraries or software development tools to compile, making it difficult to turn the source code into an executable computer program. In such cases the source code should be provided with a package that makes the source portable. While software packages that help to make source codes portable do exist (e.g., Autotools), using these packages requires non-negligible labor from the software developers. To reduce the burden on software developers who wish to share their code, we suggest that code authors can also share their binary executable files, so that readers can easily validate the results and test the computer programs without the need to struggle with source code that is difficult to compile.

Making computer programs publicly available comprises substantial advantages as described in Section 2, but there might also be some return for not sharing source code. For instance, re-implementation of a certain algorithm can be as important as the implementation used in the original paper. Making the source code publicly available can reduce the motivation to re-implement it, but sharing the source code does not limit other researchers from re-implementing the algorithm in their preferred programming language. Another downside is that in some cases authors of code might want to commercialize their work, or are subjected to the intellectual property (IP) policy of their institute according which the IP cannot be transferred. In these cases, authors can choose not to release the code under Free Software Foundation (FSF) licenses such as GPL (General Public License) and BSD (Berkeley Software Distribution), or Open Source Initiative (OSI) licenses such as AFL (Academic Free License), thus preventing for-profit organizations from using it for commercial purposes without the consent of the authors. It should be noted that once the scientific work is published, if it is not patented it is subjected to commercialization by for-profit organizations regardless of the availability of the code since readers can re-implement the code based on the methodology described in the paper, and use their implementation for commercial purposes. Another concern is code that was used for making scientific discoveries, but the code itself cannot be released for national security reasons.

An issue related to code sharing is training in software development practices. Astronomers often teach themselves coding and lack the requisite knowledge of best practices in software development. A promising approach to rectifying this state of affairs is for scientists to attend a short, intensive training class, such as the "boot camps" offered by the Software Carpentry project (http://software-carpentry.org/). The best practices recommended in the boot camps are summarized in (Wilson et al. 2012).

## 5. Conclusion

We believe that a requirement for making computer programs publicly available is in strong agreement with the actual needs of the astrophysics and astroinformatics community, and will make the work published in astronomy and astrophysics journals more useful to its readers, thus improving the impact and overall quality of the work being published. The software in a project provides the only true documentation of the computational methodologies used within a scientific paper. While many agree that scientific source code should be shared, the work of the Astrophysics Source Code Library reveals that there is a substantial gap between that agreement and the current code sharing culture and practices in the astrophysics community. Therefore, we propose that the contention that the code is part of the research should be reflected not merely by a broad agreement that computer programs should be shared, but also by making the required adjustments at the level of the formal scientific communication and defining editorial policies that will lead to discoverable, accessible source code.

Editorial policies should require that editors and reviewers of scientific papers that involve computer programs ensure that the software is shared with the scientific community when the paper is published.